\documentclass[conference, 10pt]{IEEEtran}
\usepackage{hyperref}
\usepackage{array}
\usepackage{amsmath,amsfonts}
\usepackage{algorithmic}
\usepackage{algorithm}
\usepackage{array}
\usepackage[caption=false,font=normalsize,labelfont=sf,textfont=sf]{subfig}
\usepackage{textcomp}
\usepackage{stfloats}
\usepackage{url}
\usepackage{verbatim}
\usepackage{graphicx}
\usepackage{cite}
\usepackage{amssymb}
\usepackage{color}
\usepackage{booktabs}
\usepackage{makecell}
\usepackage{pifont}
\usepackage{rotating}
\usepackage{multirow}
\usepackage{siunitx} 
\usepackage{booktabs}
\newcommand{\cmark}{\checkmark}
\begin{document}

\title{Multimodal-Wireless: A Large-Scale Dataset for Sensing and Communication\\
}
\author{\IEEEauthorblockN{Tianhao Mao$^{1}$, Le Liang$^{1,2}$, Jie Yang$^{3,4}$, Hao Ye$^{5}$, Shi Jin$^{1}$, and Geoffrey Ye Li$^{6}$}
\IEEEauthorblockA{$^1$School of Information Science and Engineering, Southeast University, Nanjing 210096, China}
\IEEEauthorblockA{$^2$Purple Mountain Laboratories, Nanjing 211111, China}
\IEEEauthorblockA{$^3$Key Laboratory of Measurement and Control of Complex Systems of Engineering, Ministry of \\Education, Southeast University, Nanjing 210096, China}
\IEEEauthorblockA{$^4$ Frontiers Science Center for Mobile Information Communication and Security, Southeast University, Nanjing 210096, China}
\IEEEauthorblockA{$^5$ Department of Electrical and Computer Engineering, University of California, Santa Cruz, CA 95064, USA}
\IEEEauthorblockA{$^6$ Department of Electrical and Electronic Engineering, Imperial College London, SW7 2BX London, U.K.}

E-mail: \{tianhao, lliang, yangjie\}@seu.edu.cn, yehao@ucsc.edu, jinshi@seu.edu.cn, geoffrey.li@imperial.ac.uk.}

\maketitle



\maketitle

\begin{abstract}
This paper presents Multimodal-Wireless, a large-scale open-source dataset for multimodal sensing and communication research. The dataset is generated through an integrated
and customizable data pipeline built upon the CARLA simulator and Sionna framework, and features high-resolution communication channel state information (CSI) fully synchronized with five other sensor modalities, namely LiDAR, RGB and depth camera, inertial measurement unit (IMU) and radar, all sampled at 100 Hz. It contains approximately 160,000 frames collected across four virtual towns, sixteen communication scenarios, and three weather conditions. This paper provides a comprehensive overview of the dataset, outlining its key features, overall framework, and technical implementation details. In addition, it explores potential research applications concerning communication and collaborative perception, exemplified by beam prediction using a multimodal large language model. The dataset is open in \url{https://le-liang.github.io/mmw/}.

\end{abstract}


\section{Introduction}

\IEEEPARstart{F}{uture} communication systems are evolving towards larger antenna arrays, higher frequency bands, and wider bandwidths. For multiple-input multiple-output (MIMO) systems, high performance relies on the precise alignment of transmit and receive beams, a challenge that necessitates context-aware communication. Machine learning has emerged as a powerful tool for this task \cite{cuiyuanhao,shibingpu,LLMoverview}, yet its potential is constrained by the lack of large-scale, comprehensive dataset.

While numerous datasets have been developed for intelligent autonomous systems \cite{opencood, dair-v2x, bostontwin, deepmimo, e-flash, viwi, DeepSense}, a critical dichotomy exists. Datasets for collaborative perception (e.g., OPV2V \cite{opencood}, DAIR-V2X \cite{dair-v2x}) excel in sensory modalities like cameras and LiDAR, whereas wireless communication datasets (e.g., BostonTwin \cite{bostontwin}, DeepMIMO \cite{deepmimo}) focus on channel modeling. The integration of these two domains remains rare. Pioneering efforts to combine them, such as E-Flash \cite{e-flash}, ViWi \cite{viwi}, and DeepSense 6G \cite{DeepSense}, lack crucial elements like adverse weather conditions, high-frequency channel state information (CSI), or complete multimodal synchronization. As summarized in Table~\ref{tab:datasets}, these gaps reveal a clear need for a unified, weather-resilient, and extensible dataset for multimodal communication research.

\newcolumntype{C}[1]{>{\centering\arraybackslash}p{#1}}
\newcolumntype{L}[1]{>{\raggedright\arraybackslash}p{#1}}

\begin{table*}[t]
\centering
\caption{Open-source real-world or simulation datasets for environmental sensing}
\label{tab:datasets}
\renewcommand{\arraystretch}{1.24} 

\begin{tabular}{| L{2.45cm} | C{0.8cm} | C{0.55cm} | C{1.42cm} | C{0.55cm} | C{0.7cm} | C{0.5cm} | C{1.85cm} | C{1cm} | C{1.55cm} | L{1.5cm} |}
\hline 
\textbf{Dataset} & \textbf{LiDAR} & \textbf{RGB} & \textbf{Depth Map} & \textbf{IMU} & \textbf{Radar} & \textbf{CSI} & \textbf{Multi-Scenario} & \textbf{Weather} & \textbf{Customizable} & \textbf{Source} \\
\hline 
OPV2V \cite{opencood}       & \cmark & \cmark &        &        &        &        &   \cmark  &        &        & Simulation \\
\hline
DAIR-V2X \cite{dair-v2x}      & \cmark & \cmark &        &        &        &        &  \cmark  &        &        & Measurement \\
\hline
BostonTwin \cite{bostontwin}  &        &        &        &        &        & \cmark &          &        & \cmark & Simulation \\
\hline
DeepMIMO \cite{deepmimo}      &        &        &        &        &        & \cmark &          &        & \cmark & Simulation \\
\hline
e-Flash \cite{e-flash}        & \cmark & \cmark &        &        &        & \cmark &          &        &        & Measurement \\
\hline
ViWi \cite{viwi}              & \cmark & \cmark & \cmark &        &        & \cmark & \cmark  &        & \cmark & Simulation \\
\hline
DeepSense 6G \cite{DeepSense} & \cmark & \cmark &        &        & \cmark &        & \cmark  &        &        & Measurement \\ 
\hline
Multimodal-Wireless                          & \cmark & \cmark & \cmark & \cmark & \cmark & \cmark &  \cmark  & \cmark & \cmark & Simulation \\ 
\hline 
\end{tabular}
\end{table*}

To address this need, we present Multimodal-Wireless, a novel dataset built on CARLA \cite{carla} and Sionna \cite{sionna} frameworks. Multimodal-Wireless uniquely provides rich, multi-path channel data fully synchronized with five other sensor modalities, namely LiDAR, RGB and depth camera, inertial measurement unit (IMU) and radar, all captured at an unprecedented 100 Hz sampling rate (10 ms resolution). This high-frequency data is critical for developing dynamic, real-time communication schemes. Moreover, the dataset incorporates diverse weather scenarios—sunny, rainy, and foggy—to enable the creation of resilient models. A key innovation is its extensibility: Researchers can easily customize scenarios and generate new, tailored datasets by modifying a configuration file, making Multimodal-Wireless a valuable research tool.
   

\section{Multimodal-Wireless: Highlights} \label{sec:framework}
Although DeepSense 6G has significantly promoted beam-related research \cite{cuiyuanhao,shibingpu}, the next generation of context-aware communication demands datasets that are more comprehensive, resilient, and flexible. Multimodal-Wireless is engineered to meet this need by overcoming the key limitations of existing resources. Its primary contributions are as follows:

\begin{itemize}
    \item \textbf{100~Hz Sampled CSI for Diverse Applications.} Instead of providing only received power, Multimodal-Wireless delivers detailed, multi-path CSI from the Sionna ray-tracer at a 100~Hz sampling rate. This channel data, synchronized with the 10~ms 5G NR frame structure \cite{5gnr}, unlocks research opportunities far beyond beam prediction, enabling the exploration of advanced physical layer and MAC layer designs. Moreover, capturing all modalities at a synchronized 100~Hz, the oversampling guarantees perfect temporal alignment between sensor streams, which is a challenge with real-world hardware.

    \item \textbf{All-Weather Resilience by Design.} Recognizing that real-world systems must operate in adverse conditions, we systematically incorporate sunny, rainy, and foggy scenarios. This allows for the study of weather-induced impairments, such as LiDAR false echos due to Mie scattering and camera degradation, which is a feature largely absent in prior datasets.

    \item \textbf{Full Flexibility and Extensibility.} Our framework breaks the mold of static datasets. Researchers can easily generate their own data by modifying a single configuration file, controlling everything from environmental parameters in CARLA to communication configurations in Sionna. This turns the dataset from a static resource into a dynamic research tool.

    \item \textbf{A Unified Resource for vehicle-to-everything (V2X) Communication and Perception.} Multimodal-Wireless provides both vehicle-to-vehicle (V2V) and vehicle-to-infrastructure (V2I) channel data alongside ground-truth 3D bounding boxes. By adding CSI into traditional collaborative perception datasets like OPV2V \cite{opencood}, this unique and dual-purpose dataset is suitable for the intersection of communication and collaborative perception.
\end{itemize}

\section{Multimodal-Wireless: General Framework}
The Multimodal-Wireless dataset is based on a V2X framework, where connected-autonomous vehicles (CAVs) and road-side units (RSUs) collaboratively gather multimodal data for collaborative perception.
Its main innovation lies in incorporating wireless channel characteristics into this V2X setup. Alongside conventional sensors (LiDAR, RGB and depth cameras, radar and IMU), both RSUs and CAVs use antenna arrays for V2X communication. Specifically, the CAVs are modeled as users and RSUs as base stations (BSs).

The dataset’s generation pipeline combines CARLA \cite{carla}, the autonomous driving simulator, Sionna \cite{sionna}, the ray-tracing engine, and Blender \cite{blender}, the physical modeling software, to ensure spatial and temporal consistency across platforms.
Blender acts as a bridge to replicate CARLA’s dynamic scenes in Sionna’s ray-tracing environment, ensuring all modalities share a unified world. The process, as illustrated in Fig. \ref{fig:workflow}, comprises the following stages:
\begin{figure*}[t]
	\centerline{\includegraphics[width=6.3in]{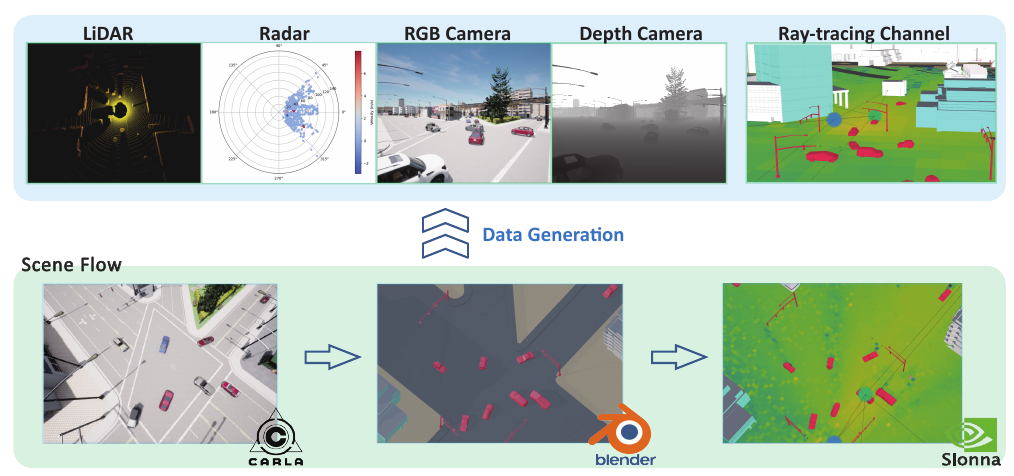}}
	\caption{Cross-platform data generation workflow for the Multimodal-Wireless dataset.}
	\label{fig:workflow}
\end{figure*}
\begin{itemize}
    \item \textbf{Scenario Execution and Data Capture in CARLA:} We first define and execute our scenario in CARLA. For each frame, five types of sensory data (LiDAR, radar, IMU, RGB and depth camera) are captured. During the simulation, the dynamic state of each frame is recorded in a configuration file, consisting of the position and rotation of CAVs and RSUs in the scenario.
    \item \textbf{Scenario Reconstruction in Blender:} The static town map from CARLA is first established in Blender as a base environment. Then, for each frame, the pose information from its configuration file is used to place and orient all dynamic actors within this environment programmatically. Each fully constituted frame is then exported as a self-contained Sionna scene.
    \item \textbf{Channel Generation in Sionna:} Finally, the exported scenes are sequentially loaded into Sionna. The transmitter and receiver locations for each link are configured based on the pose data in the configuration files. Sionna's ray-tracing engine computes the detailed channel impulse response and path parameters, which constitute the communication modality for the Multimodal-Wireless dataset.
\end{itemize}


\section{Multimodal-Wireless: Detailed Specifications} 
The integration of CARLA, Blender and Sionna facilitates the synchronization between traditional modalities and communication channel modality, as well as the spatial consistency of the CARLA world and Sionna scenes. To realize the pipeline, we need to let the scenes flow smoothly among these three platforms. In this section, we clarify the parametric and technical details in CARLA, Blender and Sionna, respectively.

\subsection{CARLA--Scenario Execution and Data Collection}
The CARLA simulator serves as the foundational environment for scenario generation within the Multimodal-Wireless framework. 
All dynamic scenarios are defined and executed in CARLA, where the five sensory modalities are first captured. In this subsection, we first detail the software architecture that facilitates the definition of scenarios, followed by a description of the scenarios already included in Multimodal-Wireless.

\begin{figure}[t]
	\centerline{\includegraphics[width=3.0in]{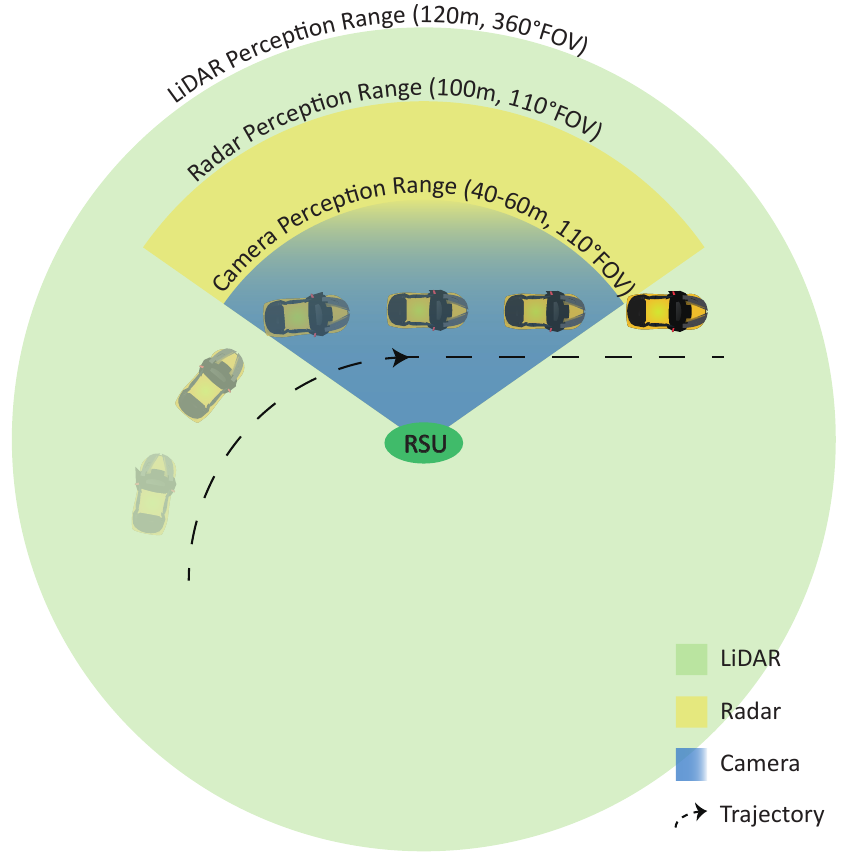}}
	\caption{Illustration of sensor perception ranges with CAV trajectory.}
	\label{fig:range}
\end{figure}

For ease of use and extensibility, our code is organized around a configuration-driven architecture. Scenarios are defined within a configuration file by three primary settings:
\begin{itemize}
    \item \textbf{Simulation Settings:} This section details the core simulation parameters. A fixed frame rate of $100$~Hz is used, with scenario durations ranging from $8$ to $13$ seconds. This configuration yields $800$ to $1300$ frames of data for each CAV per scenario. Additionally, weather-specific parameters such as precipitation, humidity, and fog density are configured to simulate sunny, rainy, and foggy conditions within the CARLA environment.
    
    \item \textbf{Scenario Settings:} This part defines the specifics of the traffic scenario, including the target town map, the number of vehicles and RSUs, and their initial placement logic. 
    We introduce two key parameters for precise agent control: $\mathbf{x}_{\mathrm{spawn}}$, a 3D coordinate around which vehicles are randomly spawned, and $\mathbf{r}$, an array of indices used to designate the \textit{n}-th closest vehicles to $\mathbf{x}_{\mathrm{spawn}}$ as the CAVs. Each scenario has three to four CAVs and one RSU.

    \item \textbf{Sensor Settings:} This part specifies the sensor suite for both CAVs and RSUs. 
    Each CAV is equipped with four RGB cameras (providing 360-degree coverage: front, back, left, right), a LiDAR, and an IMU. 
    Each RSU is equipped with an RGB camera, a depth camera, a LiDAR, and a radar. Fig.~\ref{fig:range} illustrates the perception range of each sensor, along with a schematic trajectory of a CAV. Typically, due to the limited field of view (FOV) of both camera and radar, the two sensors in the RSU only detect the CAV in part of the overall scenario, whereas the LiDAR can sense it across the entire area.
    The detailed specifications for these sensors are provided in Table~\ref{tab:lidar}.
\end{itemize}
\begin{table}[h!] 
    \centering
    \caption{Sensor specifications.}
    \label{tab:lidar}
    \renewcommand{\arraystretch}{1.3}
    \begin{tabular}{|l|l|}
        \hline 
        \textbf{Sensors} & \textbf{Attributes} \\
        \hline
        RGB Camera & $640 \times 480$ resolution, $110^{\circ}$ FOV \\
        \hline
        Depth Camera & $640 \times 480$ resolution, $110^{\circ}$ FOV \\
        \hline
        LiDAR & \makecell[l]{64 channels, $30k$ points per sample, \\ 
                            $120$ m capturing range, $-25^{\circ}$ to $2^{\circ}$ \\ 
                            vertical FOV} \\
        \hline
        IMU & \makecell[l]{Gyroscope noise: mean $0.001$ rad/s, \\
                            standard deviation (std)
                            $0.002$ rad/s, \\ accelaration noise: std $0.1$ m/s$^2$} \\
        \hline
        Radar & \makecell[l]{$2k$ points per sample, $100$ m capturing  \\ 
                            range, $30^{\circ}$ vertical FOV and $110^{\circ}$ 
                            \\ horizontal  FOV} \\
        \hline 
    \end{tabular}
\end{table}

Notably, users can define customized scenarios by modifying the configuration file to generate their own data. Existing data involves 16 distinct scenarios distributed across four diverse towns---three urban (Town03, Town05, Town10) and one rural (Town07)---each selected for its unique characteristics relevant to communication applications. We summarize all scenarios in Table~\ref{tab:scenarios}, all of which can be replayed using the provided configuration file in Multimodal-Wireless.
\begin{table*}[htbp]
\centering

\caption{MULTIMODAL-WIRELESS DATASET: TOWNS, SCENARIOS, AND FEATURES}
\vspace{-0.1cm}
\label{tab:scenarios}

\renewcommand{\arraystretch}{1.24}

\newcolumntype{C}[1]{>{\centering\arraybackslash}p{#1}}
\newcolumntype{L}[1]{>{\raggedright\arraybackslash}p{#1}}
\begin{tabular}{|L{1.1cm}|L{3.0cm}|C{0.9cm}|C{1.0cm}|C{1.0cm}|L{8cm}|}
\hline
\textbf{Town} & \textbf{Scenario} & \textbf{No. of CAVs} & \textbf{Duration (s)} & \textbf{No. of Samples} & \textbf{Feature} \\
\hline

\multirow{5}{*}{Town03} 
& Roundabout & 3 & 10 & 9,000 & \makecell[l]{Vehicles' movement in a roundabout} \\ 
\cline{2-6}
& Gas station & 3 & 8 & 7,200 & \makecell[l]{Rich reflection, and blockage under the gas station}  \\
\cline{2-6}
& T-junction with slope & 3 & 10 & 9,000 & \makecell[l]{Vertical movement due to the slope of the T-junction} \\
\cline{2-6}
& Crossroad with slope & 3 & 10 & 9,000 & \makecell[l]{Vertical movement due to the slope of the crossroad} \\
\cline{2-6}
& 5-way intersection & 3 & 11 & 9,900 & \makecell[l]{Traffic converging at a complex 5-way intersection} \\
\hline

\multirow{5}{*}{Town05} 
& Dual skybridges & 3 & 11 & 9,900 & \makecell[l]{Two connected sky-bridges that obstruct the air-to-ground LOS path} \\
\cline{2-6}
& Ring road & 3 & 13 & 11,700 & \makecell[l]{An elevated ring road
encircling part of the city} \\
\cline{2-6}
& T-junction under overpass  & 4 & 10 & 12,000 & \makecell[l]{The underside of the ringroad generating strong reflections} \\
\cline{2-6}
& CBD Crossroad & 4 & 12 & 14,400 & \makecell[l]{Buildings with glass facades with unique electromagnetic properties} \\
\cline{2-6}
& Parking lot & 3 & 11 & 9,900 & \makecell[l]{Parked cars act as scatterers} \\ 
\hline

\multirow{2}{*}{\makecell[l]{Town07\\(rural)}} 
& Single-lane road & 3 & 10 & 9,000 & \makecell[l]{Grainsilos built with bricks with unique electromagnetic properties} \\
\cline{2-6}
& Rural crossroad & 3 & 12 & 10,800 & \makecell[l]{An open-space intersection with  fewer NLOS paths } \\
\hline

\multirow{4}{*}{Town10} 
& Urban crossroad & 3 & 10 & 9,000 & \makecell[l]{Exceptionally wide lanes and more complex traffic} \\
\cline{2-6}
& Curvy road & 3 & 10 & 9,000 & \makecell[l]{An winding road with oncomming traffic encounters} \\
\cline{2-6}
& H-shaped Road & 3 & 10 & 9,000 & \makecell[l]{A road layout that facilitates U-turn maneuvers} \\
\cline{2-6}
& Wide skybridge & 3 & 11 & 9,900 & \makecell[l]{A broad overpass creating air-to-ground LOS blockage} \\
\hline

\end{tabular}
\end{table*}

\subsection{Blender--Scenario Reconstruction}
To enable ray-tracing in Sionna, the dynamic scenarios from CARLA must be reconstructed into a format compatible with electromagnetic simulation. Blender serves as the core of our scenario reconstruction and material enrichment, bridging the gap between the real-time graphics engine of CARLA and the physics simulator of Sionna. This process involves two primary stages to formulate scenes for Sionna: static environment replication and dynamic actor placement.

\begin{figure*}[t]
	\centerline{\includegraphics[width=6.6in]{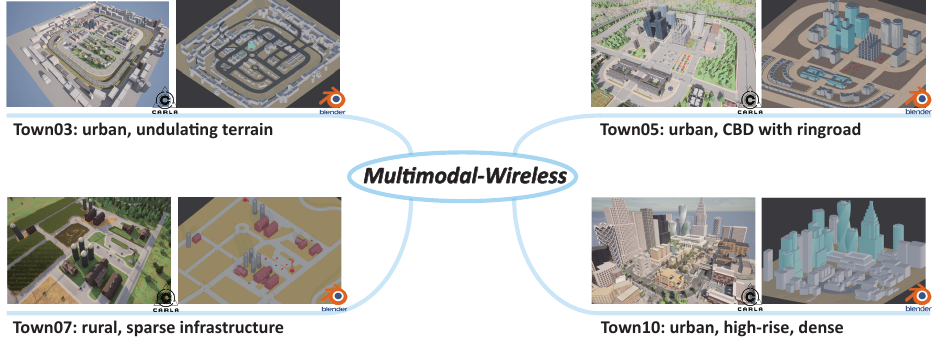}}
	\caption{A comparative overview of four standard CARLA simulation environments and their corresponding source models in Blender.}
	\label{fig:towns}
\end{figure*}

First, we address the replication of the static environment. The base geometry of the towns (roads, buildings, etc.) is exported from CARLA via Unreal Engine 4.26. However, a direct import is insufficient for accurate simulation, as standard 3D models often lack material properties crucial for ray-tracing. Therefore, we perform a critical material enrichment step in Blender. To replicate the environment's electromagnetic characteristics and ensure the subsequent ray-tracing results are physically meaningful, we allocate the materials as Table~\ref{tab:material_assignment}.
\begin{table}
\centering
\caption{Assignment of Radio Materials}
\label{tab:material_assignment}
\renewcommand{\arraystretch}{1.2}
\begin{tabular}{|m{1.5cm}|l|l|} 
\hline
\textbf{Category} & \textbf{Condition\;/\;Part} & \textbf{Assigned Material} \\
\hline
\multirow{4}{*}{Buildings} & General Facades & \texttt{itu\_marble} \\ \cline{2-3} 
                           & General Rooftops & \texttt{itu\_concrete} \\ \cline{2-3} 
                           & CBD Facades & \texttt{itu\_glass} \\ \cline{2-3} 
                           & Rural Houses & \texttt{itu\_brick} \\
\hline
\multirow{2}{*}{Ground}    & Sunny Weather & \makecell[l]{\texttt{itu\_very\_dry\_ground} \\ \texttt{itu\_medium\_dry\_ground}} \\ \cline{2-3} 
                           & Foggy\;/\;Rainy & \texttt{itu\_wet\_ground} \\
\hline
Traffic & Vehicles \& Lights & \texttt{itu\_metal} \\
\hline
\end{tabular}
\end{table}
This meticulous material assignment is essential for the simulation of the electromagnetic environment. In Fig.~\ref{fig:towns}, the high-fidelity rendering in the CARLA simulator (left) is shown alongside its mirrored 3D asset in Blender (right). The key characteristics of each environment, from rural and sparse to dense, high-rise urban settings, are also described.

Second, to reconstruct the dynamic scenarios, we develop an automated script that programmatically parses the log files for pose information per frame. For each frame, this script dynamically places the corresponding actor models into the static scene and exports the entire scene in the Sionna-compatible format. This automated workflow enables the efficient and scalable conversion of entire, complex scenarios from CARLA into a sequence of ray-tracing-ready scenes for Sionna.

\subsection{Sionna--Channel Generation}
With the dynamic scenarios fully reconstructed in a Sionna-compatible format, we proceed to the final stage of channel generation. In this stage, Sionna serves as the ray-tracing engine to generate the multi-path channel within the scenes from Blender. This subsection provides a detailed overview of the technical procedures and the channel parameters involved.

Loading the corresponding scene of each frame into Sionna, we first configure the communication links by placing antenna arrays on the RSU and CAVs. 
To ensure spatial co-location with the sensory data, these arrays are positioned at the same height as the LiDAR sensors, with their precise locations programmatically set from the log files. 
In our primary setup, we model a V2I downlink scenario, where the RSU serves as the BS and the CAVs act as mobile users. 
Notably, this framework is inherently extensible to V2V applications.

For the channel computation, we equip the RSU with a uniform linear arrays (ULA) of $N_{\mathrm{t}}$ transmit antennas and each CAV with a ULA of $N_{\mathrm{r}}$ receive antennas for 2D settings and a uniform planar array for 3D settings. 
After defining the antenna patterns and polarization, Sionna's ray-tracing engine is invoked to compute the propagation paths. 
To balance physical accuracy with computational feasibility, we collect the line-of-sight (LOS) path and all first-order reflection events. 
For each of the $M$ resulting paths, we store its fundamental physical properties: the azimuth and zenith angle-of-departure (AOD) and angle-of-arrival ($\phi_{m}^{\mathrm{t}/\mathrm{r}}$, $\theta_{m}^{\mathrm{t}/\mathrm{r}}$), the propagation delay $\tau_m$, and the matrix $\mathbf{A}_m \in \mathbb{C}^{N_{\mathrm{r}}\times N_{\mathrm{t}}}$, 
representing the complex gain of the $m$-th path at the carrier frequency $f_{\mathrm{c}}$.

A key design philosophy of Multimodal-Wireless is to decouple the raw, physics-based path data from the final frequency-domain channel realization. 
This provides users with maximum flexibility. Based on the path properties stored, we introduce a Python utility function in Sionna that allows users to generate custom single-carrier or multi-carrier (e.g., orthogonal frequency division multiplexing (OFDM)) frequency-domain channels with $\{\mathbf{A}_m\}_{m=1}^{M}$ and $\{\tau_m\}_{m=1}^{M}$. 
This is achieved by coherently summing the contributions of all paths at frequency $f_k$, according to the relationship:
\vspace{-0.4pt}\begin{equation}
\label{eq:Sionna_channel}
    \mathbf{H}(f_k) = \sum_{m=1}^{M} \mathbf{A}_{m} e^{-j2\pi f_k \tau_{m}},
\end{equation}
where $\mathbf{H}(f_k)$ is the channel frequency response at the $k$-th subcarrier. Here, $f_k$ is the baseband frequency of the $k$-th subcarrier and satisfies $f_k=(k-\frac{K+1}{2})\Delta f$ for an OFDM system with $K$ subcarriers and $\Delta f$ subcarrier spacing. Simulation parameters of the dataset are summarized in Table~\ref{tab:Sionna}, and categorized into two groups: \textbf{ray-tracing parameters}, which are set during the physics-based ray-tracing process, and \textbf{communication parameters}, which can be customized by the user when synthesizing the frequency-domain channel from the raw path data ($\{\mathbf{A}_m, \tau_m\}_{m=1}^{M}$). This mechanism grants users maximum control over the final channel realization.

\begin{table}[h!] 
    \centering
    \caption{Sionna Simulation Parameters in Multimodal-Wireless.}
    \label{tab:Sionna}
    \renewcommand{\arraystretch}{1.24}
    \begin{tabular}{|c|l|l|} 
        \hline 
        \quad & \textbf{Parameter} & \textbf{Value} \\
        \hline
        \multirow{5}{*}{\begin{turn}{90} \hspace{-1.5em} \textbf{Ray-tracing} \end{turn}} 
        & Carrier Frequency & $28$~GHz\;/\;$4.9$~GHz \\
        \cline{2-3}
        & Antenna Pattern & Dipole \\
        \cline{2-3}
        & Ray Samples Launched & $10^6$ \\
        \cline{2-3}
        & Maximum Reflection Order & $1$ \\
        \cline{2-3}
        & Polarization & Vertical \\
        \hline
        \multirow{6}{*}{\begin{turn}{90} \hspace{-1.5em} \textbf{Communication} \end{turn}} 
        & Subcarrier Spacing & $120$ kHz   \\
        \cline{2-3}
        & Number of Subcarriers & $1024$   \\
        \cline{2-3}
        & Transmit Array Size &  $1\!\times\!4/16/64/256$ (2D)\;/\;$8\!\times\!8$ (3D)\!\!\!\\
        \cline{2-3}
        & Receive Array Size & $1\!\times\!4/16$ (2D)\;/\;$8\!\times\!8$ (3D)   \\
        \cline{2-3}
        & Number of Transmit Antennas\!\!\! & $4/16/64/256$ (2D)\;/\;$64$ (3D)  \\
        \cline{2-3}
        & Number of Receive Antennas\!\!\! & $4/16$ (2D)\;/\;$64$ (3D)  \\
        \cline{2-3}
        & Frame Duration & $10$ ms  \\
        \hline 
    \end{tabular}
\end{table}

To ensure physically accurate ground reflections at 28GHz mmWave frequencies, we augment the simulation environment with three new materials based on the ITU-R Recommendation P.527-5 \cite{itu}, which are summarized in Table~\ref{tab:material}.
\begin{table}[h!]
\centering
\caption{Electromagnetic Properties of Defined Ground Materials.}

\label{tab:material}
\renewcommand{\arraystretch}{1.2}
\begin{tabular}{| p{3.5cm} | >{\centering\arraybackslash}p{2.66cm} | >{\centering\arraybackslash}p{1.4cm} |}
\hline
\textbf{\!Material Name} & \textbf{Relative Permittivity} & \textbf{Conductivity} \\
\hline
\texttt{\!itu\_very\_dry\_grnd\_28}   & 2.5 & 0.03 \\
\hline
\texttt{\!itu\_medium\_dry\_grnd\_28} & 3   & 0.4  \\
\hline
\texttt{\!itu\_wet\_grnd\_28}        & 3   & 2.5  \\
\hline
\end{tabular}
\end{table}

\section{Example Research Applications}
The rich modality of Multimodal-Wireless supports research concerning multimodal sensing and communication. For wireless communication, the inclusion of V2X CSI enables wireless interactions between users and BS, facilitating research on channel estimation, beamforming, blockage prediction, etc. For collaborative perception, the ground truth of bounding box allows for tasks such as target detection and path planning in autonomous driving.

\begin{figure}[t]
	\centerline{\includegraphics[width=3.2in]{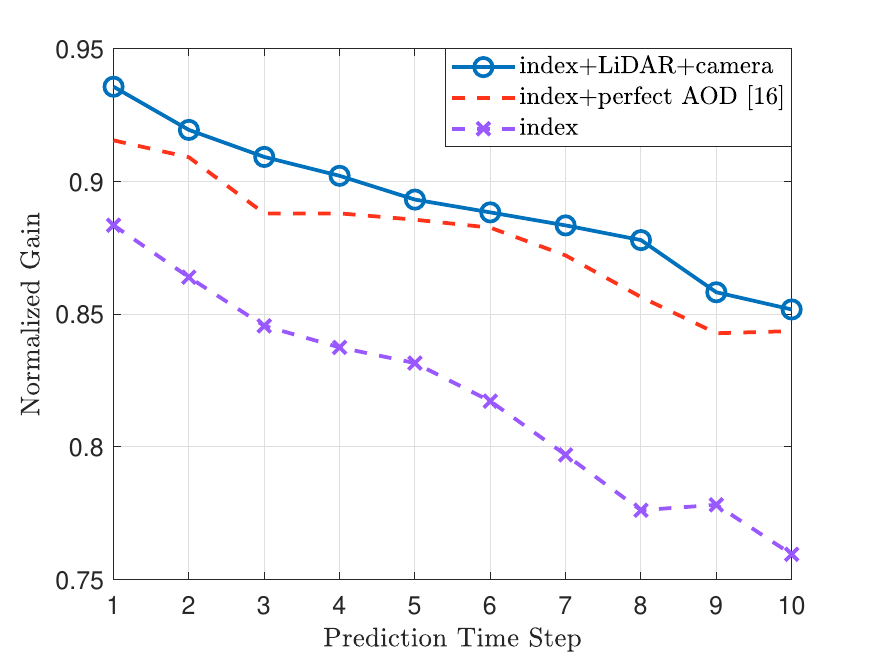}}
	\caption{Prediction performance of the proposed multimodal LLM-based method compared with the method in \cite{bpllm}.}
	\label{fig:main}
\end{figure}

\begin{figure}[t]
	\centerline{\includegraphics[width=3.2in]{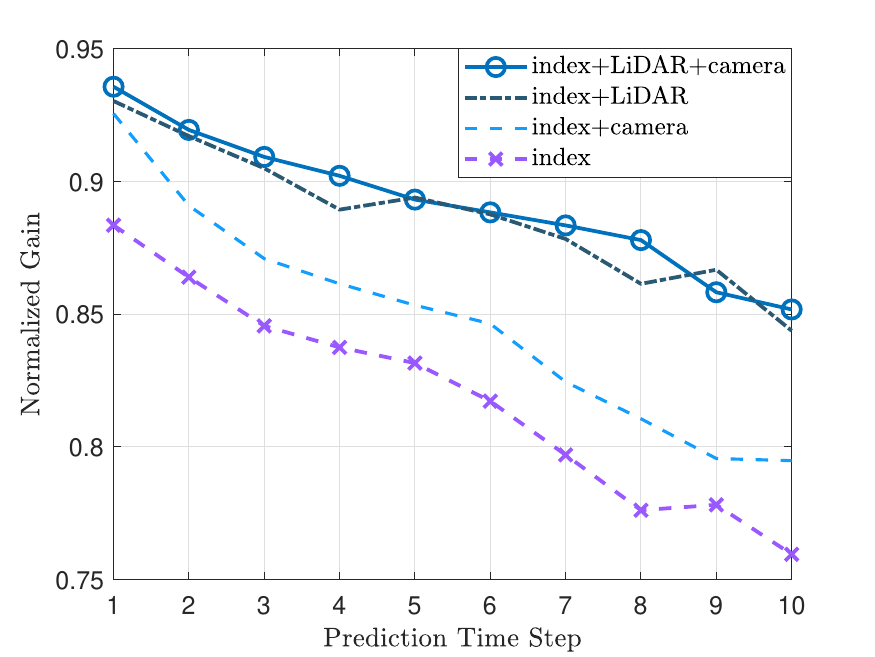}}
	\caption{Ablation study on different input modalities.}
	\label{fig:ablation}
\end{figure}

As a case study, we present multimodal large language model (LLM) based beam prediction. Specifically, we predict future beam indices of $W=10$ time steps (i.e., $100$~ms) with history beam indices, LiDAR and RGB camera data of $P=40$ time steps (i.e., $400$ ms). We utilize the data collected under the ``sunny'' weather, with a training set of 43,040 samples and a validation set of 5,380 samples constructed. We employ a pre-trained GPT-2 model as the LLM backbone. For modality-alignment, the 10~Hz RGB and LiDAR data are replicated to synchronize with the 100~Hz beam index sequence. BS applies a ULA of size $1\times64$ and user of size $1\times16$. We select the combiner $\mathbf{w}$ and the precoder $\mathbf{f}$ from a $Q$-DFT codebook, $Q$ set to 64, where the $q$-th candidate beamformer is 
\begin{equation}
    \mathbf{f}(q)=\frac{1}{\sqrt{N_t}}
    \begin{bmatrix}
        1, & e^{j2\pi q/Q}, & \cdots, & e^{j2\pi(N_t-1)q/Q}
    \end{bmatrix}^{\mathrm{T}}.
\end{equation}
We define normalized gain as performance metric by
\begin{equation} \label{eq:normalizedgain}
    G(\hat{q}_n)=\frac{|\mathbf{w}(p_n^*)^{\mathrm{H}}\mathbf{H}_n \mathbf{f}(\hat{q}_n)|^2}{|\mathbf{w}(p_n^*)^{\mathrm{H}}\mathbf{H}_n \mathbf{f}(q_n^*)|^2},
\end{equation}
where $\mathbf{H}_n$ is the channel matrix averaged over subcarriers. The optimal precoder and combiner indices, $q_n^*$ and $p_n^*$, are selected by an exhaustive search to maximize the beamforming gain as
\begin{equation} \label{eq:optimal_beam}
    (p_n^*, q_n^*) = \arg\max_{p_n, q_n \in \{0, \dots, Q-1\}} \left|\mathbf{w}(p_n)^{\mathrm{H}} \mathbf{H}_n \mathbf{f}(q_n)\right|^2.
\end{equation}
As shown in Fig.~\ref{fig:main}, by incorporating LiDAR and camera data, our multimodal LLM-based method achieves higher normalized gain than the method in \cite{bpllm} across all prediction time steps, addressing the absence of AOD on the BS side in practice. This improvement arises because historical multimodal data provide valuable environmental context, allowing the model to better capture scene dynamics such as incoming traffic jams. Fig.~\ref{fig:ablation} reveals that combining LiDAR with beam index data achieves performance close to the full-modality model. In contrast, retaining only the camera input leads to a degradation, which is owing to the limited FOV of the RGB camera that compromises the quality of the camera modality data. However, compared with the index-only baseline, the addition of the camera modality still improves performance.

\section{Conclusion} \label{sec:conclusion}
In this paper, we introduced Multimodal-Wireless, an open-sourced large-scale multimodal sensing dataset for communication. We begin with its key contributions, followed by an overview of the data generation pipeline as well as technical and parametric specifications. Finally, we discuss example applications of the dataset and demonstrate its effectiveness via a beam prediction experiment based on a multimodal LLM.

\section*{Acknowledgement}
The work was supported by the National Key R\&D Program of China under Grant 2024YFE0200700, and the National Natural Science Foundation of China (NSFC) under Grant W2421087.


\vfill

\end{document}